\documentclass[12pt]{article}
\usepackage{epsfig}
\usepackage{amsmath}
\usepackage{latexsym}
\title{Factorization in Graviton Scattering and
the ``Natural" Value of the g-factor}
\author{Barry R. Holstein\\
Department of Physics-LGRT\\
University of Massachusetts\\
Amherst, MA  01003}
\begin{document}
\begin{titlepage}
\maketitle
\begin{abstract}
The factorization property of graviton scattering amplitudes is
reviewed and shown to be valid only if the ``natural" value of the
gyromagnetic ratio $g_S=2$ is employed---independent of spin.
\end{abstract}

\end{titlepage}

\section{Introduction}

Over the years there have been a number of speculations concerning
the ``natural" value of the g-factor (gyromagnetic ratio) for a
particle of arbitrary spin.  These basically fall into two
categories.  The first is the Belinfante conjecture, which asserts
that the ``natural" value is $g_S=1/S$ for a particle of spin
$S$\cite{bel}.  This hypothesis reproduces the well-known Dirac
value---$g_{S= {1\over 2}}=2$---for a particle of spin 1/2 but
predicts smaller numbers for particles of higher spin.  A second
proposal is that the value $g_S=2$ is the ``natural" value
independent of spin\cite{tel}.  Of course, in some sense any such
speculation is somewhat metaphysical in that a ``natural" value
for the g-factor has no experimental basis for particles which
participate in the strong interactions.  One case which {\it does}
have direct empirical support is that of the charged leptons
which, carrying spin 1/2, agree with their Dirac value---$g_{\rm
Dirac}=2$---up to small electromagnetic corrections\cite{pdg}. The
only other case which has empirical support is that of the charged
$W$-boson, which for reasons discussed below, has $g_W$=2 in the
standard model\cite{ait}.  The present experimental
limits---$g=2.20\pm 0.20$\cite{pdg}---are in agreement with the
standard model prediction.

Below then in section 2 we briefly review the previous arguments
in this regard, while in section 3 we present a new argument which
favors the hypothesis $g_S=2$---the factorization of gravitational
amplitudes.  A brief concluding section follows.

\section{``Natural" g-factor}

Every student learns in his/her first quantum mechanics course
that the ``natural" value for the g-factor of a spin-1/2 particle
is its Dirac value---$g_{S={1\over 2}}=2$---and this result is
strongly confirmed experimentally in the case of the charged
leptons ($e,\mu,\tau)$ up to small electromagnetic
corrections\cite{pdg}. Of course, for particles such as the
proton, which possesses an experimental g-factor nearly three
times this value, one is not surprised because the ``natural"
value of 2 is modified by large strong interaction corrections. In
fact, this is the situation with nearly every other known
particle---strong interaction corrections obliterate any
underlying ``bare" value of the g-factor, making any direct
experimental confrontation impossible. Nevertheless, it is
intriguing to speculate theoretically what this value might be.
One of the first physicists to do so was Belinfante\cite{bel}.
Using minimal substitution he calculated directly the g-factor for
a charged particle carrying 3/2 and determined $g_{S={3\over
2}}=2/3$.  Knowing the result for unit
spin\cite{bjd}---$g_{S=1}=1$---he suggested that the ``natural"
value for particles of {\it arbitrary} spin $S$ is $g_S=1/S$. This
proposal has become known as the "Belinfante conjecture" and has
in fact been confirmed rigorously by later authors in the case
that the electromagnetic interaction is introduced via minimal
substitution\cite{oth}.

Despite this theoretical confirmation there have developed a
number of reasons to doubt the naturalness of Belinfante's
suggestion. One is the feature that besides the charged leptons,
the only other charged particle which does not have strong
interactions---the $W^\pm$-boson---does {\it not} obey this
prediction. Rather, in the standard model we have
$g_{W\pm}=2$\cite{ait}.  Since, as mentioned above, this number
has been confirmed experimentally, it is important to understand
where the difference from Belinfante's calculation comes about.

\subsection{$W^\pm$ Boson}

A neutral spin 1 field $\phi_\mu(x)$ having mass $m$ is described
by the Proca Lagrangian density, which is of the form\cite{pro}
\begin{equation}
{\cal L}(x)=-{1\over 4}U_{\mu\nu}(x)U^{\mu\nu}(x)+{1\over
2}m^2\phi_\mu(x)\phi^\mu(x)\label{eqn:la}
\end{equation}
where
\begin{equation}
U_{\mu\nu}(x)=i\partial_\mu \phi_\nu(x)-i\partial_\nu\phi_\mu(x)
\end{equation}
is the spin 1 field tensor.  If the particle has charge $e$, we
can generate a gauge-invariant form of Eq. \ref{eqn:la} by use of
the well-known minimal substitution\cite{msu}---defining
\begin{equation}
\pi_\mu=i\partial_\mu-eA_\mu(x)
\end{equation}
and
\begin{equation}
U_{\mu\nu}(x)=\pi_\mu\phi_\nu(x)-\pi_\nu \phi_\mu(x)
\end{equation}
the charged Proca Lagrangian density becomes
\begin{equation}
{\cal L}(x)=-{1\over 2}U_{\mu\nu}^\dagger(x)
U^{\mu\nu}(x)+m^2\phi_\mu^\dagger(x)\phi^\mu(x)
\end{equation}
Introducing the left-right derivative
\begin{equation}
D(x)\overleftrightarrow{\nabla} F(x)\equiv D(x)\nabla F(x)-(\nabla
D(x)) F(x)
\end{equation}
the single-photon component of the interaction can be written as
\begin{equation}
{\cal L}_{int}(x)=ieA^\mu(x)
\phi^{\alpha\dagger}(x)[\eta_{\alpha\beta}\overleftrightarrow{\nabla}_\mu
-\eta_{\beta\mu}\nabla_\alpha]\phi^\beta(x)+\eta_{\alpha\mu}(\nabla_\beta
\phi^{\alpha\dagger}(x))\phi^\beta(x)
\end{equation}
so that the on-shell matrix element of the electromagnetic current
becomes
\begin{equation}{1\over
\sqrt{4E_fE_i}}<p_f,\epsilon_B|j_\mu|p_i,\epsilon_A>=-{e\over
\sqrt{4E_fE_i}}\left[2P_\mu\epsilon_B^*\cdot\epsilon_A-
\epsilon_{A\mu}\epsilon_B^*\cdot
q+\epsilon_{B\mu}^*\epsilon_A\cdot q\right]\label{eq:br}
\end{equation}
where we have used the property
$p_f\cdot\epsilon_B^*=p_i\cdot\epsilon_A=0$ for the Proca
polarization vectors.  If we now look at the spatial piece of this
term we find
\begin{equation}
{1\over
\sqrt{4E_fE_i}}<p_f,\epsilon_B|\vec{\epsilon_\gamma}\cdot\vec{j}
|p_i,\epsilon_A>\simeq {e\over
2m}\vec{\epsilon}_\gamma\times\vec{q}\cdot\hat{\epsilon}_B^*\times\hat{\epsilon_A}
={e\over 2m}<1,m_f|\vec{S}|1,m_i>\cdot\vec{B}
\end{equation}
where we have used the result that in the Breit frame for a
nonrelativistically moving particle
\begin{equation}
i\hat{\epsilon}_B^*\times\hat{\epsilon}_A=<1,m_f|\vec{S}|1,m_i>
\end{equation}
which we recognize as representing a magnetic moment interaction
with g=1. On the other hand if we take the time component of Eq.
\ref{eq:br}, we find, again in the Breit frame and a
nonrelativistically moving system
\begin{equation}
{1\over \sqrt{4E_fE_i}}<p_f,\epsilon_B|\epsilon_{0\gamma }j_0
|p_i,\epsilon_A>\simeq -e\epsilon_{0\gamma}\left[
\epsilon_B^*\cdot\epsilon_A +{1\over 2m
}(\epsilon_{A0}\hat{\epsilon}_B^*\cdot\vec{q}-
\epsilon_{B0}^*\hat{\epsilon}_A\cdot\vec{q})\right]
\end{equation}
Using
\begin{eqnarray}
\epsilon_A^0\simeq{1\over 2m}\hat{\epsilon}_A\cdot\vec{q},\qquad
\epsilon_B^0\simeq -{1\over
2m}\hat{\epsilon}_B^*\cdot\vec{q}\nonumber\\
\epsilon_B^*\cdot\epsilon_A\simeq
-\hat{\epsilon}_B^*\cdot\hat{\epsilon}_A-{1\over
2m^2}\hat{\epsilon}_B^*\cdot\vec{q}\hat{\epsilon}_A\cdot\vec{q}
\end{eqnarray}
we observe that
\begin{equation}
{1\over \sqrt{4E_fE_i}}<p_f,\epsilon_B|\epsilon_{0\gamma }j_0
|p_i,\epsilon_A>\simeq e\epsilon_{0\gamma}
\hat{\epsilon}_B^*\cdot\hat{\epsilon}_A
\end{equation}
which is the expected electric monopole term---any electric
quadrupole contributions have cancelled\cite{yob}.  Overall then,
Eq. \ref{eq:br} corresponds to a simple E0 interaction with the
charge accompanied by an M1 interaction with g-factor unity, which
is consistent with the speculation by Belinfante that for a
particle of spin $S$, $g=1/S$\cite{bel}.

Despite this suggestively simple result, however, Eq. \ref{eqn:la}
does {\it not} correctly describe the interaction of the charged
$W$-boson field, due to the feature that the $W^\pm$ are
components of an SU(2) vector field\cite{ait}. The proper Proca
Lagrangian has the form
\begin{equation}
{\cal L}(x)=-{1\over
4}\vec{U}_{\mu\nu}^\dagger(x)\cdot\vec{U}^{\mu\nu}(x)+{1\over
2}m_W^2\vec{\phi}_\mu(x)\cdot\vec{\phi}^\mu(x)\label{eqn:ld}
\end{equation}
where the field tensor $\vec{U}_{\mu\nu}(x)$ contains an
additional term on account of gauge invariance
\begin{equation}
\vec{U}_{\mu\nu}(x)=\pi_\mu\vec{U}_\nu(x)-\pi_\nu\vec{U}_\mu(x)
-ig\vec{U}_\mu(x)\times\vec{U}_\nu(x)
\end{equation}
with $g$ being the SU(2) electroweak coupling coupling constant.
The Lagrange density Eq. \ref{eqn:ld} then contains the piece
\begin{equation}
{\cal
L}_{int}(x)=-gW^{0\mu\nu}(x)(W_\mu^{+\dagger}(x)W_\nu^+(x)-W_\mu^{-\dagger}(x)W^-_\mu)(x)
\end{equation}
among (many) others.  However, in the standard model the neutral
member of the W-triplet is a linear combination of $Z^0$ and
photon fields\cite{smb}---
\begin{equation}
W_\mu^0=\cos\theta_WZ_\mu^0+\sin\theta_WA_\mu
\end{equation}
and, since $g\sin\theta_W=e$, we have a term in the interaction
Lagrangian
\begin{equation}
{\cal
L}_{int}^{(1)}(x)=-eF_{\mu\nu}(x)(W_\mu^{+\dagger}(x)W_\nu^+(x)-W_\mu^{-\dagger}(x)W^-_\mu(x))
\end{equation}
which represents an additional interaction that must be appended
to the convention Proca result.  In the Breit frame and for a
nonrelativistically moving system we have
\begin{equation}
{1\over
\sqrt{4E_fE_i}}<p_f,\epsilon_B|\vec{\epsilon_\gamma}\cdot\vec{j}^{(1)}
|p_i,\epsilon_A>\simeq {e\over
2m_W}\vec{\epsilon}_\gamma\times\vec{q}\cdot\hat{\epsilon}_B^*\times\hat{\epsilon_A}
={e\over 2m_W}<1,m_f|\vec{S}|1,m_i>\cdot\vec{B}\label{eq:j1}
\end{equation}
and
\begin{equation}
{1\over \sqrt{4E_fE_i}}<p_f,\epsilon_B|j_0^{(1)}
|p_i,\epsilon_A>\simeq -e{1\over 2m_W
}(\epsilon_A^0\hat{\epsilon}_B^*\cdot\vec{q}-
\epsilon_{B0}^*\hat{\epsilon}_A\cdot\vec{q})=-{e\over
2m_W^2}\hat{\epsilon}_B^*\cdot\vec{q}\hat{\epsilon}_A\cdot\vec{q}\label{eq:j2}
\end{equation}
The first piece---Eq. \ref{eq:j1}---constitutes an additional
magnetic moment and modifies the W-boson g-factor from its
Belinfante value of unity to its standard model value of 2. Using
\begin{equation}
{1\over
2}(\epsilon_{Bi}^*\epsilon_{Aj}+\epsilon_{Ai}\epsilon_{Bj}^*)-{1\over
3}\delta_{ij}\hat{\epsilon}_B^*\cdot\hat{\epsilon}_A=<1,m_f|{1\over
2}(S_iS_j+S_jS_i)-{2\over 3}\delta_{ij}|1,m_i>
\end{equation}
we observe that the second component---Eq. \ref{eq:j2}---implies
the existence of a quadrupole moment of size $Q=-e/M_W^2$. Both of
these results are well known predictions of the standard model for
the charged vector bosons and the standard model prediction for
the $g$-factor is experimentally confirmed---$g_W=2.20\pm 0.20$
\cite{pdg}.

Of course, a single example does not constitute a compelling case,
but it has recently been suggested, from a number of viewpoints,
that the ``natural" value of the gyromagnetic ratio for a particle
of {\it arbitrary} spin is $g_S$=2\cite{tel}.  We shall briefly
review these arguments below and then will present a argument
which buttresses this assertion.

\subsection{GDH Sum Rule}

Perhaps the first author to suggest the importance of $g_S=2$ was
Weinberg who, in Brandeis lecture notes, examined the low energy
limit of Compton scattering\cite{wei}.  In this way he was able to
generalize the Gerasimov-Drell-Hearn (GDH) sum rule, which relates
a particle's anomalous magnetic moment to a weighted integral over
its polarized photoabsorption cross sections\cite{gdh}.  In the
case of spin 1/2 this was shown by GDH to have the form
\begin{equation}
{2\pi\alpha\over M^2}\kappa^2={1\over \pi}\int_0^\infty
{d\omega\over \omega}\Delta\sigma_s(\omega)
\end{equation}
where $\kappa$ is the nucleon anomalous magnetic moment and
$$\Delta\sigma_s(\omega)=\sigma_{3\over 2}(\omega)-
\sigma_{1\over 2}(\omega)$$ is the difference between the cross
section measured with the incoming photon and target polarizations
parallel and antiparallel.  The sum rule has been well tested and
has been shown to work in the case of the nucleon\cite{gdh}.
Weinberg demonstrated that the sum rule can be generalized to
arbitrary spin provided one defines the anomalous magnetic moment
via---
$$\vec{\mu}={e\vec{S}\over 2m}g_S(1+\kappa)$$
with $g_S=2$---independent of spin.  From the perspective of the
GDH sum rule then it is suggestive that the ``natural" value of
the g-factor is $g_S=2$\cite{wei}. However, there also exists an
argument from the realm of high energy Compton
scattering\cite{tel}.

\subsection{Compton Scattering at High Energy}

We next examine high energy Compton scattering from a basic
spin-$S$ target having mass $m$ and charge $e$, and consider the
case of spin one. As discussed above, the simple Proca interaction
for a charged spin 1 system yields the Feynman rules for photon
interactions\cite{bjd}
\begin{eqnarray}
PP\gamma: &=&-ie\left\{(p_f+p_i)_\mu
g_{\alpha\beta}-g_{\beta\mu}[gp_{f\alpha}-(g-1)p_{i\alpha}]
-g_{\alpha\mu}[gp_{i\beta}-(g-1)p_{f\beta}]\right\}\nonumber\\
PP\gamma\gamma:
&=&ie^2(2g_{\mu\nu}g_{\alpha\beta}-g_{\mu\alpha}g_{\nu\beta}-g_{\mu\beta}g_{\nu\alpha})
\end{eqnarray}
where, for generality, we have included an anomalous moment
(Pauli) interaction of the form
\begin{equation}
PP\gamma:
=-ie(g-1)F^{\mu\nu}(W^{+\dagger}_{\mu}W^+_\nu-W^{-\dagger}_\mu
W^-_\nu)
\end{equation}

\begin{figure}
\begin{center}
\epsfig{file=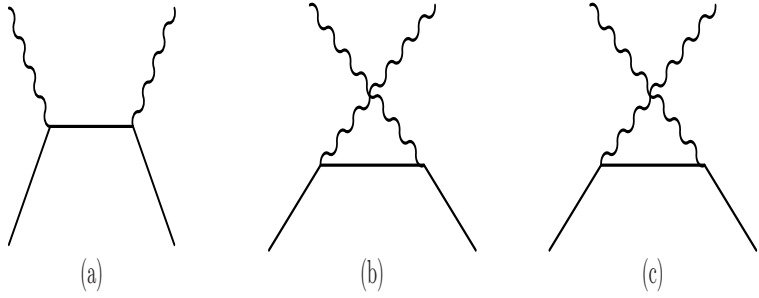,height=4cm,width=10cm} \caption{Diagrams
relevant to Compton scattering. }
\end{center}
\end{figure}

Calculation of the three lowest order diagrams shown in Figure 1
then yields the result
\begin{eqnarray}
{\rm
Amp}_{Compton}(S=1,g)&=&e^2\{2\epsilon_A\cdot\epsilon_B^*\left[{\epsilon_i\cdot
p_i\epsilon_f^*\cdot p_f\over p_i\cdot k_i}-{\epsilon_i\cdot
p_f\epsilon_f^*\cdot p_i\over p_i\cdot
k_f}-\epsilon_i\cdot\epsilon_f^*\right]\nonumber\\
&-&g\left[\epsilon_A\cdot
[\epsilon_f^*,k_f]\cdot\epsilon_B^*\left({\epsilon_i\cdot p_i\over
p_i\cdot k_i} -{\epsilon_i\cdot p_f\over p_i\cdot
k_f}\right)\right.\nonumber\\
&-&\left.\epsilon_A\cdot[\epsilon_i,k_i]\cdot\epsilon_B^*\left({\epsilon_f^*\cdot
p_f\over p_i\cdot k_i}
-{\epsilon_f^*\cdot p_i\over p_i\cdot k_f}\right)\right]\nonumber\\
&-&g^2\left[{1\over 2p_i\cdot
k_i}\epsilon_A\cdot[\epsilon_i,k_i]\cdot[\epsilon_f^*,k_f]\cdot\epsilon_B^*\right.\nonumber\\
&-&\left.{1\over 2p_i\cdot
k_f}\epsilon_A\cdot[\epsilon_f^*,k_f]\cdot[\epsilon_i,k_i]\cdot\epsilon_B^*\right]\nonumber\\
&-&{(g-2)^2\over m^2}\left[{1\over 2p_i\cdot
k_i}\epsilon_A\cdot[\epsilon_i,k_i]\cdot
p_i\epsilon_B^*\cdot[\epsilon_f^*,k_f]\cdot p_f\right.\nonumber\\
&-&\left.{1\over 2p_i\cdot
k_f}\epsilon_A\cdot[\epsilon_f^*,k_f]\cdot p_i
\epsilon_B^*\cdot[\epsilon_i,k_i]\cdot p_i\right]\}\label{eqn:co}
\end{eqnarray}
where we have defined
$$S\cdot[Q,R]\cdot T\equiv S\cdot QT\cdot R-S\cdot RT\cdot Q.$$
The interesting terms here are those on the last two lines, which
are proportional to the factor $1/m^2$.  They arise from the Born
diagrams via the $k_\alpha k_\beta/m^2$ terms of the spin-one
propagator
\begin{equation}
D_{\alpha\beta}(k)={i\over
k^2-m^2}\times\left(-g_{\alpha\beta}+{k_\alpha k_\beta\over
m^2}\right)
\end{equation}
and reveal that if we take the limit as the mass becomes small the
Compton amplitude will diverge, violating unitarity at a photon
energy $\omega_i\sim m$ {\it unless the gyromagnetic ratio has the
value $g=2$}, and this same condition can be shown to assure the
absence of $1/m^2$ terms {\it for arbitrary spin}\cite{tel}!
Again, this result is certainly suggests that the ``natural''
value of the g-factor is $g_S=2$, in agreement with the result
found from the GDH sum rule.

\subsection{Graviton Scattering and Factorization}

\begin{figure}
\begin{center}
\epsfig{file=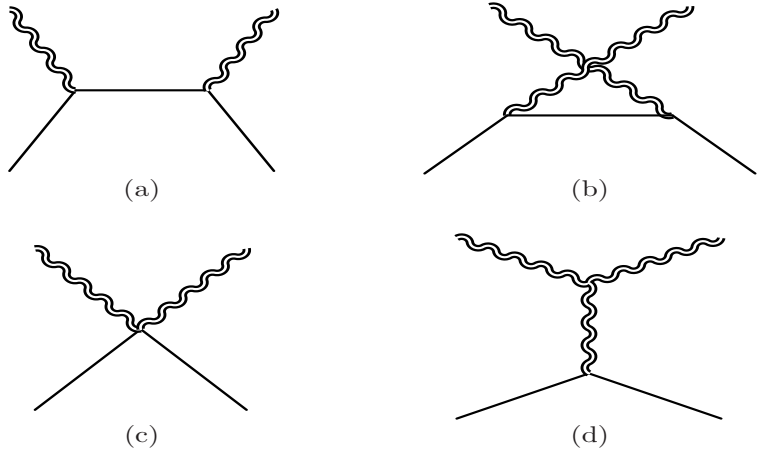,height=6cm,width=10cm} \caption{Diagrams
relevant for gravitational Compton scattering. }
\end{center}
\end{figure}

Having reviewed ``old" results\cite{asi}, we now present a new
argument which makes the case for $g=2$---factorization of
graviton scattering amplitudes.

Based on string theory arguments it has recently been pointed out
that the elastic scattering of gravitons from a ``bare" target of
arbitrary spin should factorize\cite{fac}.  This condition had
been found earlier by Song et al. from gauge theory
considerations\cite{kor}.  That is, the graviton is a particle of
spin 2 whose polarization tensor $\epsilon_{\mu\nu}$ can be
written, in harmonic gauge, which we use henceforth, as a simple
product of corresponding spin one photon polarization vectors---
$$\epsilon_{\mu\nu}^{\pm 2}=\epsilon_\mu^{\pm 1}\epsilon_\nu^{\pm 1}$$
The elastic scattering of gravitons from a target of arbitrary
spin is constructed by summing the four diagrams shown in Figure
2, consisting of two Born pieces, a seagull term, and the graviton
pole diagram.  The factorization theorem asserts that for
scattering from a target of spin $S$, the graviton scattering
amplitude from an "ideal" target particle of spin $S$ can be
written in the form
\begin{equation}
\epsilon^*_{f\alpha\beta}M_{graviton}^{\alpha\beta;\mu\nu}(S)\epsilon_{i\mu\nu}
={\kappa^2\over
8e^4}F\times[\epsilon^*_{f\alpha}A_{Compton}^{\alpha;\mu}(S)\epsilon_{i\mu}]
\times[\epsilon^*_{f\beta}
A_{Compton}^{\beta;\nu}(0)\epsilon_{i\nu}]
\end{equation}
where $M_{graviton}^{\alpha\beta;\mu\nu}(S)$ is the elastic
graviton scattering amplitude from a system of spin $S$,
$A_{Compton}^{\alpha;\mu}(S)$ is the elastic Compton amplitude
from a target of spin $S$ and charge $e$, F is the kinematic
factor
\begin{equation}
F={p_1\cdot k_1p_1\cdot k_2\over k_1\cdot k_2}
\end{equation}
and $\kappa^2=32\pi G$ is the gravitional coupling  This is a
remarkable result and dramatically simplifies the evaluation of
graviton scattering. In the case of spin 0 the calculation is
fairly straightforward. The four diagrams which must be included
in order to satisfy gauge invariance are shown in Figure 2. From
the Klein-Gordon Lagrangian
\begin{equation}
{\cal L}_{S=0}={1\over
2}\left(\partial_\mu\phi\partial^\mu\phi-m^2\phi^2\right)
\end{equation}
we find the one- and two-graviton vertices\cite{boh}
\begin{eqnarray}
\tau_{\alpha\beta}(p,p')&=&{-i\kappa\over
2}\left(p_\alpha{p'}_\beta+{p'}_\alpha
p_\beta-\eta_{\alpha\beta}(p\cdot p'-m^2)\right)\nonumber\\
\tau_{\alpha\beta,\gamma\delta}(p,p')&=&i\kappa^2\left[
I_{\alpha\beta,\rho\xi}I^\xi{}_{\sigma,\gamma\delta}
\left(p^\rho{p'}^\sigma+{p'}^\rho p^\sigma\right)\right.\nonumber\\
&-&\left.{1\over
2}\left(\eta_{\alpha\beta}I_{\rho\sigma,\gamma\delta}
+\eta_{\gamma\delta}I_{\rho\sigma,\alpha\beta}\right){p'}^\rho
p^\sigma\right.\nonumber\\
&-&\left.{1\over 2}\left(I_{\alpha\beta,\gamma\delta} -{1\over
2}\eta_{\alpha\beta}\eta_{\gamma\delta}\right) \left(p\cdot
p'-m^2\right)\right]
\end{eqnarray}
while the triple graviton vertex has the form\cite{don}
\begin{eqnarray}
\tau^{\mu\nu}_{\alpha\beta,\gamma\delta}(k,q)&=&{i\kappa\over
2}\left\{ P_{\alpha\beta,\gamma\delta} \left[k^\mu k^\nu+(k-q)^\mu
(k-q)^\nu+q^\mu q^\nu-{3\over
2}\eta^{\mu\nu}q^2\right]\right.\nonumber\\
&+&\left.2q_\lambda
q_\sigma\left[I^{\lambda\sigma,}{}_{\alpha\beta}I^{\mu\nu,}
{}_{\gamma\delta}+I^{\lambda\sigma,}{}_{\gamma\delta}I^{\mu\nu,}
{}_{\alpha\beta}-I^{\lambda\mu,}{}_{\alpha\beta}I^{\sigma\nu,}
{}_{\gamma\delta}-I^{\sigma\nu,}{}_{\alpha\beta}I^{\lambda\mu,}
{}_{\gamma\delta}\right]\right.\nonumber\\
&+&\left.[q_\lambda
q^\mu(\eta_{\alpha\beta}I^{\lambda\nu,}{}_{\gamma\delta}
+\eta_{\gamma\delta}I^{\lambda\nu,}{}_{\alpha\beta})+ q_\lambda
q^\nu(\eta_{\alpha\beta}I^{\lambda\mu,}{}_{\gamma\delta}
+\eta_{\gamma\delta}I^{\lambda\mu,}{}_{\alpha\beta})\right.\nonumber\\
&-&\left.q^2(\eta_{\alpha\beta}I^{\mu\nu,}{}_{\gamma\delta}+\eta_{\gamma\delta}
I^{\mu\nu,}{}_{\alpha\beta})-\eta^{\mu\nu}q^\lambda
q^\sigma(\eta_{\alpha\beta}
I_{\gamma\delta,\lambda\sigma}+\eta_{\gamma\delta}
I_{\alpha\beta,\lambda\sigma})]\right.\nonumber\\
&+&\left.[2q^\lambda(I^{\sigma\nu,}{}_{\alpha\beta}
I_{\gamma\delta,\lambda\sigma}(k-q)^\mu
+I^{\sigma\mu,}{}_{\alpha\beta}I_{\gamma\delta,\lambda\sigma}(k-q)^\nu\right.\nonumber\\
&-&\left.I^{\sigma\nu,}{}_{\gamma\delta}I_{\alpha\beta,\lambda\sigma}k^\mu-
I^{\sigma\mu,}{}_{\gamma\delta}I_{\alpha\beta,\lambda\sigma}k^\nu)\right.\nonumber\\
&+&\left.q^2(I^{\sigma\mu,}{}_{\alpha\beta}I_{\gamma\delta,\sigma}{}^\nu+
I_{\alpha\beta,\sigma}{}^\nu
I^{\sigma\mu,}{}_{\gamma\delta})+\eta^{\mu\nu}q^\lambda q_\sigma
(I_{\alpha\beta,\lambda\rho}I^{\rho\sigma,}{}_{\gamma\delta}+
I_{\gamma\delta,\lambda\rho}I^{\rho\sigma,}{}_{\alpha\beta})]\right.\nonumber\\
&+&\left.[(k^2+(k-q)^2)\left(I^{\sigma\mu,}{}_{\alpha\beta}I_{\gamma\delta,\sigma}{}^\nu
+I^{\sigma\nu,}{}_{\alpha\beta}I_{\gamma\delta,\sigma}{}^\mu-{1\over
2}\eta^{\mu\nu}P_{\alpha\beta,\gamma\delta}\right)\right.\nonumber\\
&-&\left.(k^2\eta_{\gamma\delta}I^{\mu\nu,}{}_{\alpha\beta}+(k-q)^2\eta_{\alpha\beta}
I^{\mu\nu,}{}_{\gamma\delta})]\right\}
\end{eqnarray}
where we have defined
\begin{equation}
I_{\alpha\beta,\mu\nu}={1\over
2}(\eta_{\alpha\mu}\eta_{\beta\nu}+\eta_{\alpha\nu}\eta_{\beta\mu})
\end{equation}
and
\begin{equation}
P_{\alpha\beta,\mu\nu}=I_{\alpha\beta,\mu\nu}-{1\over
2}\eta_{\alpha\beta}\eta_{\mu\nu}
\end{equation}
The other component which we require is the graviton propagator,
which has the form in harmonic gauge
\begin{equation}
D_{\alpha\beta;\gamma\delta}(q)={iP_{\alpha\beta;\gamma\delta}\over
q^2}
\end{equation}
It is now straightforward (though tedious) to evaluate the four
diagrams, yielding
\newpage

\begin{center}
{\bf Graviton Scattering: Spin 0}
\end{center}

\begin{eqnarray}
{\rm Born-a}:&& {\rm Amp}_a(S=0)=-\kappa^2{(\epsilon_i\cdot
p_i)^2(\epsilon_f^*\cdot
p_f)^2\over p_i\cdot k_i}\nonumber\\
{\rm Born-b}:&& {\rm Amp}_b(S=0)=\kappa^2{(\epsilon_f^*\cdot
p_i)^2(\epsilon_i\cdot
p_f)^2\over p_i\cdot k_f}\nonumber\\
{\rm Seagull}:&&{\rm
Amp}_c(S=0)=\kappa^2\left[\epsilon_f^*\cdot\epsilon_i(\epsilon_i\cdot
p_i\epsilon_f^*\cdot p_f+\epsilon_i\cdot p_f\epsilon_f^*\cdot
p_i)-{1\over 2}k_i\cdot
k_f(\epsilon_f^*\cdot\epsilon_i)^2\right]\nonumber\\
{\rm g-pole}:&&{\rm Amp}_d(S=0)={\kappa^2\over 2k_i\cdot k_f}
\left[\epsilon_f^*\cdot p_f\epsilon_f^*\cdot
p_i(\epsilon_i\cdot(p_i-p_f))^2\right.\nonumber\\
&+&\left.\epsilon_i\cdot p_i\epsilon_i\cdot
p_f(\epsilon_f^*\cdot(p_i-p_f))^2\right.\nonumber\\
&+&\left.\epsilon_i\cdot(p_i-p_f)\epsilon_f^*\cdot(p_f-p_i)(\epsilon_f^*\cdot
p_f\epsilon_i\cdot p_i+\epsilon_f^*\cdot p_i\epsilon_i\cdot
p_f)\right.\nonumber\\
&-&\left.\epsilon_f^*\cdot\epsilon_i\left(\epsilon_i\cdot(p_i-p_f)\epsilon_f^*\cdot
(p_f-p_i)(p_i\cdot p_f-m^2)\right.\right.\nonumber\\
&+&\left.\left.k_i\cdot k_f(\epsilon_f^*\cdot p_f\epsilon_i\cdot
p_i+\epsilon_f^*\cdot p_i\epsilon_i\cdot p_f)
+\epsilon_i\cdot(p_i-p_f)(\epsilon_f^*\cdot p_fp_i\cdot
k_f+\epsilon_f^*\cdot p_ip_f\cdot k_f)\right.\right.\nonumber\\
&+&\left.\left.\epsilon_f^*\cdot(p_f-p_i)(\epsilon_i\cdot
p_ip_f\cdot
k_i+\epsilon_i\cdot p_fp_i\cdot k_i)\right)\right.\nonumber\\
&+&\left.(\epsilon_f^*\cdot\epsilon_i)^2\left(p_i\cdot k_ip_f\cdot
k_i+p_i\cdot k_fp_f\cdot k_f-{1\over 2}(p_i\cdot k_ip_f\cdot
k_f+p_i\cdot k_fp_f\cdot k_i)\right.\right.\nonumber\\
&+&\left.\left.{3\over 2}k_i\cdot k_f(p_i\cdot
p_f-m^2)\right)\right]
\end{eqnarray}
and when combined, one verifies that
\begin{eqnarray}
{\rm Amp}_a(S=0)&+&{\rm Amp}_b(S=0)+{\rm Amp}_c(S=0)+{\rm
Amp}_d(S=0)\nonumber\\
&=&{\kappa^2\over 8e^4}F\times[{\rm Amp}_{Compton}(S=0)]^2
\end{eqnarray}
where
\begin{equation}
{\rm Amp}_{Compton}(S=0)=2e^2\left[{\epsilon_{i}\cdot
p_i\epsilon_f^*\cdot p_f\over p_i\cdot k_i}-{\epsilon_i\cdot
p_f\epsilon_f^*\cdot p_i\over p_i\cdot
k_f}-\epsilon_f^*\cdot\epsilon_i\right]
\end{equation}
is the Compton scattering amplitude from a spinless target.
\newpage

Similarly in the case of spin 1/2 we must use the vierbein
${e_a}^\mu$ in order to define the Dirac Lagrangian\cite{boh}
\begin{equation}
{\cal L}_{S={1\over 2}}=\bar{\psi}(i\gamma^a{e_a}^\mu D_\mu-m)\psi
\end{equation}
where the covariant derivative $D_\mu$ is given by
\begin{equation}
D_\mu=\partial_\mu+{i\over 4}\sigma^{ab}\omega_{\mu ab}
\end{equation}
with
\begin{equation}
\omega_{\mu ab}={1\over 2}{e_a}^\nu(\partial_\mu
e_{b\nu}-\partial_\nu e_{b\mu})-{1\over 2}{e_b}^\nu(\partial_\mu
e_{a\nu}-\partial_\nu e_{a\mu})+{1\over
2}{e_a}^\rho{e_b}^\sigma(\partial_\sigma e_{c\rho}-\partial_\rho
e_{c\sigma}){e_\mu}^c
\end{equation}
The resulting one and two graviton vertices are then:

\begin{eqnarray}
\tau_{\alpha\beta}(p,p')&=&{-i\kappa\over 2}\left[{1\over
4}(\gamma_\alpha(p+p')_\beta+\gamma_\beta(p+p')_\alpha)-\eta_{\alpha\beta}({1\over
2}(\not\!\!{p}+\not\!\!{p}')-m)\right]\nonumber\\
\tau_{\alpha\beta,\gamma\delta}(p,p')&=&i\kappa^2\left\{-{1\over
2}({1\over
2}(\not\!\!{p}+\not\!\!{p}')-m)P_{\alpha\beta,\gamma\delta}\right.\nonumber\\
&-&\left.{1\over
16}[\eta_{\alpha\beta}(\gamma_\gamma(p+p')_\delta+\gamma_\delta(p+p')_\gamma)
\right.\nonumber\\
&+&\left.\eta_{\gamma\delta}(\gamma_\alpha
(p+p')_\beta+\gamma_\beta(p+p')_\alpha)]\right.\nonumber\\
&+&\left.{3\over
16}(p+p')^{\epsilon}\gamma^{\xi}(I_{\xi\phi,\alpha\beta}{I^{\phi}}_{\epsilon,\gamma\delta}
+I_{\xi\phi,\gamma\delta}{I^{\phi}}_{\epsilon,\alpha\beta})\right.\nonumber\\
&-&\left.{i\over 16}\epsilon^{\rho\sigma\eta\lambda}\gamma_\lambda
\gamma_5({I_{\alpha\beta,\eta}}^\nu
I_{\gamma\delta,\sigma\nu}{k'}_\rho-{I_{\gamma\delta,\eta}}^\nu
I_{\alpha\beta,\sigma\nu}k_\rho)\right\}
\end{eqnarray}
The calculation is somewhat more challenging than in the case of
spin 0 because of the Dirac algebra, but is still straightforward.
Indeed, evaluating the same diagrams we find the individual
contributions:
\newpage
\begin{center}
{\bf Graviton Scattering: Spin ${1\over 2}$}
\end{center}

\begin{eqnarray}
{\rm Born-a}:&&{\rm Amp}_a(S={1\over
2})=-\kappa^2{\epsilon_f^*\cdot p_f\epsilon_i\cdot p_i\over
8p_i\cdot
k_i}\bar{u}(p_f)[\not\!{\epsilon_f}^*(\not\!{p}_i+\not\!{k}_i+m)\not\!{\epsilon}_i]u(p_i)\nonumber\\
{\rm Born-b}:&&{\rm Amp}_b(S={1\over
2})=\kappa^2{\epsilon_f^*\cdot p_i\epsilon_i\cdot p_f\over
8p_i\cdot
k_f}\bar{u}(p_f)[\not\!{\epsilon_i}(\not\!{p}_i-\not\!{k}_f+m)\not\!{\epsilon}_f^*]u(p_i)\nonumber\\
{\rm Seagull}:&& {\rm Amp}_c(S={1\over 2
})=\kappa^2\bar{u}(p_f)\left[{3\over
16}\epsilon_f^*\cdot\epsilon_i(\not\!{\epsilon_i}\epsilon_f^*\cdot(p_i+p_f)
+\not\!{\epsilon}_f^*\epsilon_i\cdot(p_i+p_f))\right.\nonumber\\
&-&\left.{i\over
16}\epsilon_f^*\cdot\epsilon_i\epsilon^{\rho\sigma\eta\lambda}
\gamma_\lambda\gamma_5(\epsilon_{i\eta}\epsilon_{f\sigma}^*k_{f\rho}-
\epsilon_{f\eta}^*\epsilon_{i\sigma}k_{i\rho})\right]u(p_i)\nonumber\\
{\rm g-pole}:&&{\rm Amp}_d(S={1\over 2})={\kappa^2\over 16k_i\cdot
k_f}\bar{u}(p_f)\nonumber\\
&\times&\left[\not\!{k}_i\epsilon^*_f\cdot\epsilon_i
(-2\epsilon_i\cdot(p_i+p_f)\epsilon_f^*\cdot
k_i+\epsilon_f^*\cdot\epsilon_ik_i\cdot(p_i+p_f))\right.\nonumber\\
&+&\left.\not\!{k}_f\epsilon^*_f\cdot\epsilon_i
(-2\epsilon_f^*\cdot(p_i+p_f)\epsilon_i\cdot
k_f+\epsilon_f^*\cdot\epsilon_ik_i\cdot(p_i+p_f))\right.\nonumber\\
&+&\left.4\not\!{\epsilon}_i[\epsilon_f^*\cdot k_i(\epsilon_i\cdot
p_i\epsilon_f^*\cdot p_f-\epsilon_f^*\cdot p_i\epsilon_i\cdot
p_f)+\epsilon_f^*\cdot\epsilon_i(\epsilon_f^*\cdot p_i p_f\cdot
k_i-\epsilon_f^*\cdot p_fk_i\cdot p_i)]\right.\nonumber\\
&+&\left.4\not\!{\epsilon}_f^*[\epsilon_i\cdot k_f(\epsilon_i\cdot
p_i\epsilon_f^*\cdot p_f-\epsilon_f^*\cdot p_i\epsilon_i\cdot p_f
)+\epsilon_f^*\cdot\epsilon_i(\epsilon_i\cdot p_fk_f\cdot
p_i-\epsilon_i\cdot p_ik_f\cdot p_f)] \right]\nonumber\\
&\times&u(p_i)
\end{eqnarray}
Combining these terms, we reproduce again the factorization
condition, but now in the form
\begin{eqnarray}
{\rm Amp}_a(S={1\over 2})&+&{\rm Amp}_b(S={1\over 2})+{\rm
Amp}_c(S={1\over 2})+{\rm Amp}_d(S={1\over 2})\nonumber\\
&=&{\kappa^2\over 8e^4}F*[{\rm Amp}_{Compton}(S=0)]*[{\rm
Amp}_{Compton}(S={1\over 2})]\nonumber\\
\quad
\end{eqnarray}
where
\begin{equation}
{\rm Amp}_{Compton}(S={1\over 2})=e^2\bar{u}(p_f)
\left[{\not\!{\epsilon}_f^*(\not\!{p}_i+\not\!{k}_i+m)\not\!{\epsilon}_i\over
2p_i\cdot
k_i}-{\not\!{\epsilon}_i(\not\!{p}_f-\not\!{k}_i+m)\not\!{\epsilon}_f^*\over
2p_f\cdot k_i}\right]u(p_i)
\end{equation}
is the spin 1/2 Compton scattering amplitude.
\newpage
Thus far we have verified factorization for spin 0 and spin 1/2,
but we have learned nothing about conditions on the g-factor for
particles of higher spin.  This situation changes when we move to
the case of spin 1, for which we use the Proca equation\cite{pro}
discussed above. The one- and two-graviton vertices are then found
to be
\begin{eqnarray}
\tau^{(1)}_{\beta,\alpha,\mu,\nu}(p_1,p_2)&=&i{\kappa\over
2}\left\{(p_{1\mu}p_{2\nu}+p_{1\nu}p_{2\mu})\eta_{\alpha\beta}\right.\nonumber\\
&-&\left.p_{1\beta}(p_{2\mu}\eta_{\nu\alpha}+p_{2\nu}\eta_{\alpha\mu})\right.\nonumber\\
&-&\left.p_{2\alpha}(p_{1\mu}\eta_{\nu\beta}+p_{1\nu}\eta_{\beta\mu})\right.\nonumber\\
&+&\left.(p_1\cdot
p_2-m^2)(\eta_{\mu\alpha}\eta_{\nu\beta}+\eta_{\mu\beta}\eta_{\nu\alpha})\right.\nonumber\\
&-&\left.\eta_{\mu\nu}[(p_1\cdot
p_2-m^2)\eta_{\alpha\beta}-p_{1\beta}p_{2\alpha}\right]\}
\nonumber\\
\tau^{(2)}_{\beta,\alpha,\mu,\nu,\rho,\sigma}(p_1,p_2)&=&-i{\kappa^2\over
4}\left\{[p_{1\beta}p_{2\alpha} -
          \eta_{\alpha\beta}(p_1\cdot p_2 - m^2)]
      (\eta_{\mu\rho}\eta_{\nu\sigma}+
          \eta_{\mu\sigma}\eta_{\nu\rho} -
          \eta_{\mu\nu}\eta_{\rho\sigma})\right.\nonumber\\
          &+&\left.
    \eta_{\mu\rho}[\eta_{\alpha\beta}(p_{1\nu}p_{2\sigma} +
                p_{1\sigma}p_{2\nu}) -
          \eta_{\alpha\nu}p_{1\beta}p_{2\sigma}-
          \eta_{\beta\nu}p_{1\sigma}p_{2\alpha}\right.\nonumber\\
          &-&\left.
          \eta_{\beta\sigma}p_{1\nu}p_{2\alpha} -
          \eta_{\alpha\sigma}p_{1\beta}
            p_{2\nu} + (p_1\cdot p_2 -
                m^2)(\eta_{\alpha\nu}\eta_{\beta\sigma} +
                \eta_{\alpha\sigma}\eta_{\beta\nu})]\right.\nonumber\\
                &+&\left.
    \eta_{\mu\sigma}[\eta_{\alpha\beta}(p_{1\nu}p_{2\rho} +
                p_{1\rho}p_{2\nu}) -
          \eta_{\alpha\nu}p_{1\beta}p_{2\rho} -
          \eta_{\beta\nu}p_{1\rho}p_{2\alpha}\right.\nonumber\\
          &-&\left.
          \eta_{\beta\rho}p_{1\nu}p_{2\alpha}-
          \eta_{\alpha\rho}p_{1\beta}
            p_{2\nu} + (p_1\cdot p_2 -
                m^2)\eta_{\alpha\nu}\eta_{\beta\rho} +
                \eta_{\alpha\rho}\eta_{\beta\nu})]\right.\nonumber\\
                &+&\left.
    \eta_{\nu\rho}[\eta_{\alpha\beta}(p_{1\mu}p_{2\sigma} +
                p_{1\sigma}p_{2\mu})
                -\eta_{\alpha\mu}p_{1\beta}p_{2\sigma} -
          \eta_{\beta\mu}p_{1\sigma}p_{2\alpha}\right.\nonumber\\
          &-&\left.\eta_{\beta\sigma}p_{1\mu}p_{2\alpha}
          -\eta_{\alpha\sigma}p_{1\beta}
            p_{2\mu} + (p_1\cdot p_2 -
                m^2)(\eta_{\alpha\mu}\eta_{\beta\sigma} +
                \eta_{\alpha\sigma}\eta_{\beta\mu})]\right.\nonumber\\
                &+&\left.
    \eta_{\nu\sigma}[\eta_{\alpha\beta}(p_{1\mu}p_{2\rho} +
                p_{1\rho}p_{2\mu}) -
          \eta_{\alpha\mu}p_{1\beta}p_{2\rho} -
          \eta_{\beta\mu}p_{1\rho}p_{2\alpha}\right.\nonumber\\
          &-&\left.\eta_{\beta\rho}p_{1\mu}p_{2\alpha}-\eta_{\alpha\rho}p_{1\beta}
            p_{2\mu} + (p_1\cdot p_2 -
                m^2)(\eta_{\alpha\mu}\eta_{\beta\rho} +
                \eta_{\alpha\rho}\eta_{\beta\mu})]\right.\nonumber\\
                &-&\left.
    \eta_{\mu\nu}[\eta_{\alpha\beta}(p_{1\rho}p_{2\sigma} +
                p_{1\sigma}p_{2\rho}) -
          \eta_{\alpha\rho}p_{1\beta}p_{2\sigma} -
          \eta_{\beta\rho}p_{1\sigma}p_{2\alpha}\right.\nonumber\\
          &-&\left.\eta_{\beta\sigma}p_{1\rho}p_{2\alpha}-
          \eta_{\alpha\sigma}p_{1\beta}p_{2\rho} + (p_1\cdot p_2 -
                m^2)(\eta_{\alpha\rho}\eta_{\beta\sigma} +
                \eta_{\beta\rho}\eta_{\alpha\sigma})]\right.\nonumber\\
                &-&\left.
    \eta_{\rho\sigma}[\eta_{\alpha\beta}(p_{1\mu}p_{2\nu} +
                p_{1\nu}p_{2\mu}) -
          \eta_{\alpha\mu}p_{1\beta}p_{2\nu} -
          \eta_{\beta\mu}p_{1\nu}p_{2\alpha}\right.\nonumber\\
          &-&\left.
          \eta_{\beta\nu}p_{1\mu}p_{2\alpha} -
          \eta_{\alpha\nu}p_{1\beta}
            p_{2\mu} + (p_1\cdot p_2 -
                m^2)(\eta_{\alpha\mu}\eta_{\beta\nu} +
                \eta_{\beta\mu}\eta_{\alpha\nu})]\right.\nonumber\\
                 &+&\left.
    (\eta_{\alpha\rho}p_{1\mu} -
          \eta_{\alpha\mu}p_{1\rho})(\eta_{\beta\sigma}
            p_{2\nu} - \eta_{\beta\mu}p_{2\sigma})\right.\nonumber\\
            &+&\left.
    (\eta_{\alpha\sigma}p_{1\nu} -
          \eta_{\alpha\nu}p_{1\sigma})\eta_{\beta\rho}
            p_{2\mu} - \eta_{\beta\mu}p_{2\rho})\right.\nonumber\\
            &+&\left.
    (\eta_{\alpha\sigma}p_{1\mu} -
          \eta_{\alpha\mu}p_{1\sigma})(\eta_{\beta\rho}
          p_{2\nu} - \eta_{\beta\nu}p_{2\rho})\right.\nonumber\\
          &+&\left.
    (\eta_{\alpha\rho}p_{1\nu} -
          \eta_{\alpha\nu}p_{1\rho})(\eta_{\beta\sigma}
            p_{2\mu} - \eta_{\beta\mu}p_{2\sigma})\right\}
\end{eqnarray}
and the individual contributions from the four diagrams are
somewhat more complex:
\newpage
\begin{center}
{\bf Graviton Scattering: Spin 1}
\end{center}

\begin{eqnarray}
{\rm Born-a}:&& {\rm Amp}_a(S=1)=\kappa^2{1\over 2p_i\cdot
k_i}[(\epsilon_i\cdot p_i)^2 (\epsilon_f^*\cdot
p_f)^2\epsilon_A\cdot\epsilon_B^*\nonumber\\
&-&(\epsilon_f^*\cdot p_f)^2\epsilon_i\cdot p_i (\epsilon_A\cdot
k_i\epsilon_B^*\cdot \epsilon_i+
\epsilon_A\cdot\epsilon_i\epsilon_B^*\cdot p_i)\nonumber\\
&-&(\epsilon_i\cdot p_i)^2\epsilon_f^*\cdot p_f
(\epsilon_B^*\cdot\epsilon_f^*\epsilon_A\cdot
p_f+\epsilon_B^*\cdot
k_f\epsilon_A\cdot\epsilon_f^*)\nonumber\\
&+&\epsilon_i\cdot p_i\epsilon_f^*\cdot p_f\epsilon_i\cdot
p_f\epsilon_A\cdot
k_i\epsilon_B^*\cdot\epsilon_f^*+\epsilon_i\cdot
p_i\epsilon_f^*\cdot p_f\epsilon_f^*\cdot p_i\epsilon_A\cdot
\epsilon_i\epsilon_B^*\cdot k_f\nonumber\\
&+&(\epsilon_f^*\cdot
p_f)^2\epsilon_B^*\cdot\epsilon_i\epsilon_A\cdot\epsilon_i
p_i\cdot k_i+(\epsilon_i\cdot
p_i)^2\epsilon_B^*\cdot\epsilon_f^*\epsilon_A\cdot\epsilon_f^*
p_f\cdot k_f\nonumber\\
&+&\epsilon_i\cdot p_i\epsilon_f^*\cdot p_f(\epsilon_A\cdot
k_i\epsilon_B^*\cdot k_f\epsilon_i\cdot
\epsilon_f^*+\epsilon_B^*\cdot\epsilon_f^*\epsilon_A\cdot\epsilon_ip_i\cdot
p_f)\nonumber\\
&-&\epsilon_i\cdot p_i\epsilon_f^*\cdot p_i\epsilon_B^*\cdot
\epsilon_f^*\epsilon_A\cdot\epsilon_ip_f\cdot
k_f-\epsilon_f^*\cdot p_f\epsilon_i\cdot
p_f\epsilon_A\cdot\epsilon_i\epsilon_B^*\cdot\epsilon_f^*p_i\cdot
k_i\nonumber\\
&-&\epsilon_i\cdot p_i\epsilon_A\cdot k_i\epsilon_B^*\cdot
\epsilon_f^*\epsilon_f^*\cdot\epsilon_ip_f\cdot
k_f-\epsilon_f^*\cdot p_f\epsilon_B^*\cdot
k_f\epsilon_A\cdot\epsilon_i\epsilon_i\cdot\epsilon_f^*p_i\cdot
k_i\nonumber\\
&+&\epsilon_A\cdot\epsilon_i\epsilon_B^*\cdot\epsilon_f^*p_i\cdot
k_ip_f\cdot
k_f\epsilon_i\cdot\epsilon_f^*-m^2\epsilon_B^*\cdot\epsilon_f^*
\epsilon_A\cdot\epsilon_i\epsilon_f^*\cdot
p_f\epsilon_i\cdot p_i]\nonumber\\
{\rm Born-b}:&& {\rm Amp}_b(S=1)=-\kappa^2{1\over 2p_i\cdot
k_f}[(\epsilon_f^*\cdot p_i)^2 (\epsilon_i\cdot
p_f)^2\epsilon_A\cdot\epsilon_B^*\nonumber\\
&+&(\epsilon_i\cdot p_f)^2\epsilon_f^*\cdot p_i
(\epsilon_A\cdot k_f\epsilon_B^*\cdot\epsilon_f^*
-\epsilon_A\cdot\epsilon_f^*\epsilon_B^*\cdot p_i)\nonumber\\
&+&(\epsilon_f^*\cdot p_i)^2\epsilon_i\cdot p_f (\epsilon_B^*\cdot
k_i\epsilon_A\cdot\epsilon_i-\epsilon_B^*\cdot\epsilon_i\epsilon_A\cdot p_f)\nonumber\\
&-&\epsilon_f^*\cdot p_i\epsilon_i\cdot p_f\epsilon_f^*\cdot
p_f\epsilon_A\cdot
k_f\epsilon_B^*\cdot\epsilon_i-\epsilon_f^*\cdot
p_i\epsilon_i\cdot p_f\epsilon_i\cdot p_i\epsilon_A\cdot
\epsilon_f^*\epsilon_B^*\cdot k_i\nonumber\\
&-&(\epsilon_i\cdot
p_f)^2\epsilon_B^*\cdot\epsilon_f^*\epsilon_A\cdot\epsilon_f^*
p_i\cdot k_f-(\epsilon_f^*\cdot
p_i)^2\epsilon_B^*\cdot\epsilon_i\epsilon_A\cdot\epsilon_i
p_f\cdot k_i\nonumber\\
&+&\epsilon_f^*\cdot p_i\epsilon_i\cdot p_f(\epsilon_A\cdot
k_f\epsilon_B^*\cdot k_i\epsilon_i\cdot
\epsilon_f^*+\epsilon_B^*\cdot\epsilon_i\epsilon_A\cdot\epsilon_f^*p_i\cdot
p_f)\nonumber\\
&+&\epsilon_f^*\cdot p_i\epsilon_i\cdot p_i\epsilon_B^*\cdot
\epsilon_i\epsilon_A\cdot\epsilon_f^*p_f\cdot k_i+\epsilon_i\cdot
p_f\epsilon_f^*\cdot
p_f\epsilon_A\cdot\epsilon_f^*\epsilon_B^*\cdot\epsilon_ip_i\cdot
k_f\nonumber\\
&-&\epsilon_f^*\cdot p_i\epsilon_A\cdot k_f\epsilon_B^*\cdot
\epsilon_i\epsilon_i\cdot\epsilon_f^*p_f\cdot k_i-\epsilon_i\cdot
p_f\epsilon_B^*\cdot
k_i\epsilon_A\cdot\epsilon_f^*\epsilon_f^*\cdot\epsilon_ip_i\cdot
k_f\nonumber\\
&+&\epsilon_A\cdot\epsilon_f^*\epsilon_B^*\cdot\epsilon_ip_i\cdot
k_fp_f\cdot
k_i\epsilon_i\cdot\epsilon_f^*-m^2\epsilon_B^*\cdot\epsilon_i
\epsilon_A\cdot\epsilon_f^*\epsilon_i\cdot
p_f\epsilon_f^*\cdot p_i]\nonumber\\
{\rm seagull}:&&{\rm Amp}_c(S=1)=-{\kappa^2\over 4}
[(\epsilon_i\cdot\epsilon_f^*)^2(m^2-p_i\cdot
p_f)\epsilon_A\cdot\epsilon_B^*+\epsilon_A\cdot
p_f\epsilon_B^*\cdot
p_i(\epsilon_i\cdot\epsilon_f^*)^2\nonumber\\
&+&\epsilon_i\cdot p_i\epsilon_f^*\cdot
p_f(2\epsilon_i\cdot\epsilon_f^*\epsilon_A\cdot\epsilon_B^*-
2\epsilon_A\cdot\epsilon_2\epsilon_B^*\cdot\epsilon_1)\nonumber\\
&+&\epsilon_i\cdot p_f\epsilon_f^*\cdot
p_i(2\epsilon_i\cdot\epsilon_f^*\epsilon_A\cdot\epsilon_B^*-
2\epsilon_A\cdot\epsilon_i\epsilon_B^*\cdot\epsilon_f^*)\nonumber\\
&+&2\epsilon_i\cdot p_i\epsilon_1\cdot
p_f\epsilon_A\cdot\epsilon_f^*\epsilon_B^*\cdot\epsilon_f^*+2\epsilon_f^*\cdot
p_f\epsilon_f^*\cdot
p_i\epsilon_A\cdot\epsilon_i\epsilon_B^*\cdot\epsilon_i\nonumber\\
&-&2\epsilon_i\cdot p_i\epsilon_i\cdot\epsilon_f^*\epsilon_A\cdot
p_f\epsilon_B^*\cdot\epsilon_f^*-2\epsilon_f^*\cdot
p_f\epsilon_i\cdot\epsilon_f^*\epsilon_A\cdot\epsilon_i\epsilon_f^*\cdot
p_i\nonumber\\
&-&2\epsilon_i\cdot
p_f\epsilon_i\cdot\epsilon_f^*\epsilon_A\cdot\epsilon_f^*\epsilon_B^*\cdot
p_i-2\epsilon_f^*\cdot
p_i\epsilon_i\cdot\epsilon_f^*\epsilon_B^*\cdot\epsilon_i\epsilon_A\cdot
p_f\nonumber\\
&-&2(m^2-p_f\cdot
p_i)\epsilon_i\cdot\epsilon_f^*(\epsilon_A\cdot\epsilon_i\epsilon_B^*\cdot\epsilon_f^*
+\epsilon_A\cdot\epsilon_f^*\epsilon_B^*\cdot\epsilon_i)]
\end{eqnarray}
and finally the (lengthy) graviton pole contribution is
\begin{eqnarray}
 {\rm
g-pole}:&&{\rm Amp}_d(S=1)=-{\kappa^2\over 16k_i\cdot
k_f}\{\epsilon_B^*\cdot\epsilon_A[(\epsilon_i\cdot\epsilon_f^*)^2(4k_i\cdot
p_ip_f\cdot
k_i+4k_f\cdot p_ik_f\cdot p_f\nonumber\\
&-&2(p_i\cdot k_ip_f\cdot k_f+p_f\cdot k_ip_i\cdot k_f)+6p_i\cdot
p_fk_i\cdot k_f)\nonumber\\
&+&4((\epsilon_i\cdot k_f)^2\epsilon_f^*\cdot p_f\epsilon_f^*\cdot
p_i+(\epsilon_f^*\cdot k_i)^2\epsilon_i\cdot p_i\epsilon_i\cdot
p_f\nonumber\\
&+&\epsilon_i\cdot k_f\epsilon_f^*\cdot k_i(\epsilon_i\cdot
p_i\epsilon_f^*\cdot p_f+\epsilon_i\cdot p_f\epsilon_f^*\cdot
p_i))\nonumber\\
&-&4\epsilon_i\cdot\epsilon_f^*(\epsilon_i\cdot
k_f(\epsilon_f^*\cdot
p_i p_f\cdot k_f+\epsilon_f^*\cdot p_fk_f\cdot p_i)\nonumber\\
&+&\epsilon_f^*\cdot k_i(\epsilon_i\cdot p_ip_f\cdot
k_i+\epsilon_i\cdot p_fp_i\cdot k_i))\nonumber\\
&-&4k_i\cdot k_f\epsilon_i\cdot\epsilon_f^*(\epsilon_i\cdot
p_i\epsilon_f^*\cdot p_f+\epsilon_i\cdot p_f\epsilon_f^*\cdot
p_i)-4p_i\cdot p_f\epsilon_i\cdot\epsilon_f^*\epsilon_i\cdot
k_f\epsilon_f^*\cdot k_i]\nonumber\\
&-&(p_i\cdot p_f\epsilon_B^*\cdot\epsilon_A-\epsilon_B^*\cdot
p_i\epsilon_A\cdot p_f)[10(\epsilon_i\cdot\epsilon_f^*)^2k_i\cdot
k_f+4\epsilon_i\cdot\epsilon_f^*\epsilon_i\cdot
k_f\epsilon_f^*\cdot
k_i\nonumber\\
&-&4(\epsilon_i\cdot\epsilon_f^*)^2k_i\cdot
k_f-8\epsilon_i\cdot\epsilon_f^*\epsilon_i\cdot
k_f\epsilon_f^*\cdot
k_i]\nonumber\\
&+&(p_i\cdot
p_f-m^2)[(\epsilon_i\cdot\epsilon_f^*)^2(4\epsilon_A\cdot
k_i\epsilon_B^*\cdot k_i+4\epsilon_A\cdot k_f\epsilon_B^*\cdot
k_f\nonumber\\
&-&2(\epsilon_A\cdot k_i\epsilon_B^*\cdot k_f+\epsilon_A\cdot
k_f\epsilon_B^*\cdot k_i)+6\epsilon_B^*\cdot\epsilon_Ak_i\cdot
k_f)\nonumber\\
&+&4[(\epsilon_i\cdot
k_f)^2\epsilon_A\cdot\epsilon_f^*\epsilon_B^*\cdot\epsilon_f^*+(\epsilon_f^*\cdot
k_i)^2\epsilon_A\cdot\epsilon_i\epsilon_B^*\cdot\epsilon_i\nonumber\\
&+&\epsilon_i\cdot k_f\epsilon_f^*\cdot
k_f(\epsilon_A\cdot\epsilon_i\epsilon_B^*\cdot\epsilon_f^*
+\epsilon_A\cdot\epsilon_f^*\epsilon_B^*\cdot\epsilon_i)]\nonumber\\
&-&4\epsilon_i\cdot\epsilon_f^*[\epsilon_i\cdot
k_f(\epsilon_A\cdot\epsilon_f^*\epsilon_B^*\cdot
k_f+\epsilon_B^*\cdot\epsilon_f^*\epsilon_A\cdot k_f)\nonumber\\
&+&\epsilon_f^*\cdot
k_i(\epsilon_A\cdot\epsilon_i\epsilon_B^*\cdot
k_i+\epsilon_B^*\cdot\epsilon_i\epsilon_A\cdot k_i)\nonumber\\
&+&k_i\cdot
k_f(\epsilon_A\cdot\epsilon_i\epsilon_B^*\cdot\epsilon_f^*+
\epsilon_B^*\cdot\epsilon_i\epsilon_A\cdot\epsilon_f^*)+\epsilon_A\cdot\epsilon_B^*\epsilon_i\cdot
k_f\epsilon_f^*\cdot k_i]]\nonumber\\
&-&2\epsilon_A\cdot
p_f[(\epsilon_f^*\cdot\epsilon_i)^2[2\epsilon_B^*\cdot k_ip_i\cdot
k_i+2\epsilon_B^*\cdot k_fp_i\cdot k_f+3\epsilon_B^*\cdot
p_ik_i\cdot k_f\nonumber\\
&-&(\epsilon_B^*\cdot k_ip_i\cdot k_f+\epsilon_B^*\cdot
k_fp_i\cdot k_i)]\nonumber\\
&+&2(\epsilon_i\cdot
k_f)^2\epsilon_B^*\cdot\epsilon_f^*\epsilon_f^*\cdot
p_i+2(\epsilon_f^*\cdot
k_i)^2\epsilon_B^*\cdot\epsilon_i\epsilon_i\cdot p_i\nonumber\\
&+&2\epsilon_i\cdot k_f\epsilon_f^*\cdot
k_i(\epsilon_B^*\cdot\epsilon_i\epsilon_f^*\cdot
p_i+\epsilon_i\cdot
p_i\epsilon_B^*\cdot\epsilon_f^*)\nonumber\\
&-&2\epsilon_i\cdot\epsilon_f^*[\epsilon_i\cdot
k_f(\epsilon_B^*\cdot\epsilon_f^*p_i\cdot k_f+\epsilon_f^*\cdot
p_i\epsilon_B^*\cdot k_f)\nonumber\\
&+&\epsilon_f^*\cdot k_i(\epsilon_B^*\cdot\epsilon_ip_i\cdot
k_i+\epsilon_B^*\cdot k_i\epsilon_i\cdot p_i)]\nonumber\\
&-&2k_i\cdot
k_f\epsilon_i\cdot\epsilon_f^*(\epsilon_B^*\cdot\epsilon_i\epsilon_f^*\cdot
p_i+\epsilon_B^*\cdot\epsilon_f^*\epsilon_i\cdot p_i)
-2\epsilon_B^*\cdot p_i\epsilon_i\cdot\epsilon_f^*\epsilon_i
\cdot k_f\epsilon_f^*\cdot k_i]\nonumber\\
&-&2\epsilon_B^*\cdot
p_i[(\epsilon_f^*\cdot\epsilon_i)^2[2\epsilon_A\cdot k_ip_f\cdot
k_i+2\epsilon_A\cdot k_fp_f\cdot k_f+3\epsilon_A\cdot p_fk_i\cdot
k_f\nonumber\\
&-&(\epsilon_A\cdot k_ip_f\cdot k_f+\epsilon_A\cdot k_fp_f\cdot
k_f)]\nonumber\\
&+&2(\epsilon_i\cdot
k_f)^2\epsilon_A\cdot\epsilon_f^*\epsilon_f^*\cdot
p_f+2(\epsilon_f^*\cdot
k_i)^2\epsilon_A\cdot\epsilon_i\epsilon_i\cdot p_f\nonumber\\
&+&2\epsilon_i\cdot k_f\epsilon_f^*\cdot
k_i(\epsilon_A\cdot\epsilon_i\epsilon_f^*\cdot p_f+\epsilon_i\cdot
p_f\epsilon_A\cdot\epsilon_f^*\nonumber\\
&-&2\epsilon_i\cdot\epsilon_f^*[\epsilon_i\cdot
k_f(\epsilon_A\cdot\epsilon_f^*p_f\cdot k_f+\epsilon_f^*\cdot
p_f\epsilon_A\cdot k_f)\nonumber\\
&+&\epsilon_f^*\cdot k_i(\epsilon_A\cdot\epsilon_ip_f\cdot
k_i+\epsilon_A\cdot k_i\epsilon_i\cdot p_f)]\nonumber\\
&-&2k_i\cdot
k_f\epsilon_i\cdot\epsilon_f^*(\epsilon_A\cdot\epsilon_i\epsilon_f^*\cdot
p_f+\epsilon_A\cdot\epsilon_f^*\epsilon_i\cdot p_f)
-2\epsilon_A\cdot p_f\epsilon_i\cdot\epsilon_f^*\epsilon_i\cdot
k_f\epsilon_f^*\cdot k_i]\}\nonumber\\
\quad
\end{eqnarray}
Nevertheless, when combined (after considerable effort) one finds
once again the factorization condition to be valid, this time in
the form
\begin{eqnarray} {\rm Amp}_a(S=1)&+&{\rm Amp}_b(S=1)+{\rm
Amp}_c(S=1)+{\rm Amp}_d(S=1)\nonumber\\
&=&{\kappa^2\over 8e^4}F*[{\rm Amp}_{Compton}(S=0)]*[{\rm
Amp}_{Compton}(S=1,g=2)]\nonumber\\
\quad\label{eqn:gr}
\end{eqnarray}
where ${\rm Amp}_{Compton}(S=1,g=2)$ is the Compton scattering
amplitude quoted earlier in Eq. \ref{eqn:co} {\it with the
g-factor set equal to 2}.

From the gravitational side of Eq, \ref{eqn:gr}, the full graviton
scattering amplitude might naively be expected to contain terms
proportional to $1/m^2$ from the Born diagrams and the piece of
the spin 1 propagator proportional to $1/m^2$. However, this does
{\it not} occur, as can be seen from the half-off-shell form of
the single graviton coupling given above
\begin{eqnarray}
<p_2,\lambda|T_{\mu\nu}|p_1,\epsilon_A>&=&\epsilon_{A\lambda}(p_{2\mu}p_{1\nu}+p_{1\nu}p_{2\mu})
+g_{\mu\nu} p_{1\lambda}\epsilon_A\cdot p_2\nonumber\\
&-&p_{1\lambda}(p_{2\mu}\epsilon_{A\nu}+p_{2\nu}\epsilon_{A\mu})-\epsilon_A\cdot
p_2(p_{1\mu}g_{\lambda\nu}+p_{1\nu}g{\lambda\mu})\nonumber\\
&+&(p_2\cdot
p_1-m^2)(g_{\lambda\mu}\epsilon_{A\nu}+\epsilon_{A\mu}g_{\lambda\nu}-g_{\mu\nu}\epsilon_{A\lambda})
\end{eqnarray}
Setting $p_2=p_1+k$ and contracting with the intermediate state
momentum $(p_1+k)^\lambda$ we find a result proportional to
$m^2$---
\begin{equation}
(p_1+k)^\lambda
<p_1+k_1,\lambda|T_{\mu\nu}|p_1,\epsilon_A>=m^2(g_{\mu\nu}\epsilon_A\cdot
k_1-\epsilon_{A\mu}(p_1+k_1)_\nu-\epsilon_{A\nu}(p_1+k_1)_\mu)
\end{equation}
This term cancels the factor $1/m^2$ from the spin one propagator
so that no term proportional to $1/m^2$ survives in the Born
amplitude and this vanishing of terms which diverge as
$m\rightarrow 0$ can be shown to be a general property regardless
of the spin of the target.  If we acknowledge the validity of the
factorization result, then the vanishing of $1/m^2$ terms in the
gravitational amplitude can only result from the vanishing of such
terms in the corresponding Compton amplitude, which we have
already argued occurs only if the value $g=2$ is chosen, so from a
new standpoint----factorization of graviton scattering
amplitudes---we see again that the ``natural'' value for the
g-factor is $g_S=2$.

\section{Conclusions}

Above we have examined the question of the ``natural" value for
the g-factor of a particle of spin S.  Although the simple minimal
substitution gives rise to the Belinfante conjecture
$g_S=1/S$\cite{bel}, we pointed out that more recent studies have
suggested a correctness of a universal
value---$g_S=2$---independent of spin. We first pointed out that
this arises from the well known features:

\begin{itemize}
\item [i)] the standard model g-factor of the charged W-boson is 2

\item [ii)] the GDH sum rule provides a measure of the quantity $(g_S-2)^2$
in the case of arbitrary spin\cite{gdh}.  If we use this sum rule
to {\it define} the anomalous magnetic moment then clearly the
``natural" value for the gyromagnetic ratio is $g_S=2$\cite{wei}

\item [iii)] in high energy Compton scattering from a target of
arbitrary spin the choice of a gyromagnetic ratio different from 2
leads to terms which are divergent in the small mass
limit\cite{tel} and which violate unitarity at photon energies
$\omega\sim m$.
\end{itemize}

We then presented a new argument for the correctness of this
assertion---factorization in graviton scattering.  Gauge
invariance and string theory arguments make the case that the
graviton scattering amplitude for a ``bare" target having spin $S$
should factor into pieces proportional to the product of the
Compton scattering amplitude for spin 0 times the Compton
scattering amplitude for spin S times a universal kinematic
factor.  Since the graviton scattering amplitude does not contain
terms involving the inverse mass squared of the target particle,
the same must be true for the Compton amplitudes, but this is true
only if the gyromagnetic ratio has the value 2.

Again we emphasize that there is little experimental content in
this prediction---except for the well-verified cases of the
charged $e,\mu,\tau$ leptons and the charged W boson, all of which
carry $g\simeq 2$ in the standard model. Nevertheless, the
question of the existence of a ``natural" value for the g-factor
is an intriguing one, to which we have provided new input.

\begin{center}
{\bf Acknowledgement}
\end{center}

This work was supported in part by the National Science Foundation
under award PHY-02-42801.  Useful conversations with John Donoghue
and Andreas Ross are gratefully acknowledged, as is the
hospitality of Prof. A. Faessler and the theoretical physics group
from the University of T\"{u}bingen, where this paper was
finished.

\end{document}